\documentclass[a4paper,12pt]{article}

\usepackage{amsmath}
\usepackage{amssymb,epsfig,psfig}

\usepackage{stmaryrd,wasysym,upgreek,latexsym}
\usepackage{lscape}
\usepackage[vflt]{floatflt}
\usepackage{afterpage}

\newtheorem{lemma}{Lemma}[section]
\newtheorem{theorem}{Theorem}[section]

\newtheorem{proposition}{Proposition}[section]

\newcommand{\qed}{{\hfill $\Box$}}

\def\NN{{\mathbb N}}

\def\NN{{\mathbb N}}

\newcommand{\beq}{\begin{equation}}
\newcommand{\eeq}{\end{equation}}
\newcommand{\beqa}{\begin{eqnarray}}
\newcommand{\eeqa}{\end{eqnarray}}
\newcommand{\noi}{\noindent}
\newcommand{\om}{\Omega}
\newcommand{\tom}{\tilde \Omega}
\newcommand{\e}{\varepsilon}

\newcommand{\resetequ}{\setcounter{equation}{0}}

\begin{document}

\title{Parametric representation of ``critical'' noncommutative QFT models}
\author{
Vincent Rivasseau and Adrian Tanas\u{a}\footnote{e-mail:
  Vincent.Rivasseau@th.u-psud.fr, adrian.tanasa@ens-lyon.org}\\
Laboratoire de Physique Th\'eorique, \\
b\^at. 210, CNRS UMR 8627,\\
Universit\'e Paris XI, 91405, Orsay Cedex, France
}
\maketitle

\begin{abstract}
We extend the parametric representation of renormalizable non commutative quantum field theories
to a class of theories which we call ``critical",
because their power counting is definitely more difficult to obtain.This class of theories is important 
since it includes gauge theories, which should be relevant for the quantum Hall effect.
\end{abstract}

\section{Introduction}
\resetequ

Quantum field theories on a non-commutative space-time or NCQFT \cite{DN} deserve 
a systematic investigation. They are intermediate structures
between ordinary quantum field theory on commutative space time
and string theories \cite{CDS}\cite{SW}. They can also be better adapted
than ordinary quantum field theory to the description of physical phenomena with 
non-local effective interactions, such as physics 
in presence of a strong background field, for example 
the quantum Hall effect
\cite{Suss}\cite{Poly}\cite{HellRaams}, and perhaps also the confinement. 

In the labyrinth of all possible Lagrangians and geometries, we propose to use 
renormalizability as an Ariane thread. Indeed renormalizable theories are the
ones who survive under renormalization group flows, hence should be 
considered the generic building blocks of physics.

Following the Grosse-Wulkenhaar breakthrough \cite{GW1}\cite{GW2}
and subsequent work
\cite{RVW}\cite{GMRV}\cite{V}, we have now a fairly good 
understanding of a first class of renormalizable NCQFT's on the simplest non commutative 
geometry, the Moyal space. These models fall into two broad categories, 
depending on their propagators:
\begin{itemize}
\item the ordinary models, such as the initial $\phi^4_4$ model
  \cite{GW1}\cite{GW2} with harmonic potential, for
which the propagator $C(x,y)$ decays both as $x-y$ and $x+y$ tend to infinity, as shown in
the ordinary Mehler kernel representation and

\item the so-called critical models whose propagators involve covariant derivatives 
in a constant external field. The ``orientable'' Gross-Neveu model in two
dimensions \cite{GMRV}\cite{V}
and the LSZ model in 4 dimensions \cite{LSZ} fall into this category. The propagator $C(x,y)$ 
in this case only decays as $x-y$ tends to infinity, and simply oscillates 
when $x+y$ tend to infinity. This second class of models is therefore 
harder to study but is the relevant one
for the quantum Hall effect and for gauge theories.
\end{itemize}

An important technical tool in ordinary QFT is the parametric representation.
It is the most condensed form of perturbation  theory, since both position and momenta have
been integrated out. It leads to the correct basic objects at the core 
of QFT and renormalization theory, namely trees. It displays both  explicit positivity
and a kind of "democracy" between these trees: indeed the various
trees all contribute to the topological polynomial of a graph with 
the same positive coefficient, as shown in (\ref{s1}). This is nothing but the old "tree matrix theorem" 
of XIXth century electric circuits adapted to Feynman graphs \cite{A}.
Finally parametric representation displays dimension of space time as an explicit 
parameter, hence it is the natural frame to define dimensional
regularization and renormalization, which 
respects the symmetries of gauge theories.

The parametric representation for ordinary renormalizable NCQFT's was computed  in 
\cite{GR}. It no longer involves ordinary polynomials in the Schwinger parameters
but new hyperbolic polynomials.  They contain richer information
than in ordinary commutative field theory, since they are based on ribbon graphs, 
and distinguish topological invariants such as the genus of the surface on 
which these graphs live. The basic objects in these polynomials are
"admissible subgraphs" which are  
more general than trees; among these subgraphs the leading terms which govern power counting
are "hypertrees" which are the disjoint union of a 
tree in the direct graph and a tree in the dual graph. Again there is
positivity and ``democracy'' 
between them. We think these new combinatorial objects will probably 
stand at the core of the (yet to be developed) non perturbative or "constructive" 
theory of NCQFT's.

In this paper we generalize the work of \cite{GR} to the more difficult second class
of renormalizable NCQFTs, namely the critical ones. The basic objects (the hypertrees) 
and the positivity theorems remain essentially the same, but the identification of the 
leading terms and the "democracy" theorem between them is much more involved. 
We rely partly on \cite{V}, in which the key difficulty 
was to check independence between the direct space oscillations coming from
the vertices and from 
the critical propagators. This independence
implied renormalizability of the orientable Gross-Neveu model.
Our more precise method uses a kind of "fourth Filk move"
inspired by \cite{V} and \cite{GR}.

This paper is organized as follows.  In the next section we briefly recall the
parametric representation for commutative QFT and we present the
noncommutative model as well as our conventions. The
third section computes the first polynomial and its ultraviolet
 leading terms. We state here our main result, Theorem \ref{main}, which  
sets an upper bound on the Feynman amplitudes. Moreover, exact power
 counting as function of the graph genus 
follows directly from this Theorem. This is an improvement with respect to 
 \cite{V}, where only weaker bounds, sufficient just for renormalizability, 
 were established. 


The fourth section  analyses then the second polynomial, the noncommutative analog
of the Symanzik polynomial \eqref{s2}. It allows us to recover also the
proper power counting dependence in the number of broken faces. Finally, in the
last section we present some explicit polynomials for different types of Feynman graphs.

\section{Parametric Representation; the Noncommutative Model}
\resetequ

\subsection{Parametric Representation for Commutative QFT}

Let us give here the results of the parametric representation for commutative
QFT (one can see for example \cite{itz} or \cite{riv} for further details).
The amplitude of a Feynman graph writes
\beqa \label{as} 
{\cal A} (p) = \delta(\sum p)\int_0^{\infty} 
\frac{e^{- V(p,\alpha)/U (\alpha) }}{U (\alpha)^{2}} 
\prod_{\ell=1}^L  ( e^{-m^2 \alpha_\ell} d\alpha_\ell )\ .
\eeqa
\noi
where $L$ is the number of internal lines of the graph and $U$ and $V$ are
polynomials of the parameters $\alpha_\ell$ ($\ell=1,\ldots, L$) associated 
to each internal line. These so called ``topological" or
``Symanzik" polynomials have the
explicit expressions:
\beqa
\label{s1}
U &=& \sum_{\cal T} \prod_{l \not \in {\cal T}} \alpha_\ell \ , 
\eeqa
\beqa
\label{s2}
V &= &\sum_{{\cal T}_2} \prod_{l \not \in {\cal T}_2} \alpha_\ell  (\sum_{i \in
  E({\cal T}_2)} p_i)^2 \ , 
\eeqa
\noi
where $\cal T$ is a (spanning) tree of the graph and ${\cal T}_2$ is a
$2-$tree, {\it i. e.} a tree minus one of its lines. 

\subsection{The Noncommutative Model}

For simplicity we treat in this paper the LSZ model in $4$ dimensions, but the
extension to the Gross-Neveu model is straightforwrd.
We place ourselves in a Moyal space of dimension $4$
\beqa
\label{2D}
[x^\mu, x^\nu]=i \Theta^{\mu \nu},
\eeqa
\noi
where the the matrix $\Theta$ is
\beqa
\label{theta}
\Theta = \begin{pmatrix} \Theta_2 & 0 \\ 0 & \Theta_2 \end{pmatrix},\ \
\Theta_2 = \begin{pmatrix} 0 & -\theta \\ \theta & 0 \end{pmatrix}.
\eeqa
\noi
The Lagrangian is 
\beqa
\label{actiunea}
{\cal L}=\frac 12 (\partial_\mu \Phi + \Omega x_\mu\Phi)
(\partial^\mu \bar \Phi + \Omega x^\mu\bar \Phi)
 + \frac{1}{4!}\bar \Phi\star\Phi\star\bar \Phi\star\Phi.
\eeqa
\noi
where the Euclidean metric is used and $\star$ is the Moyal product.
For such a model, the propagator between two  points $x$ and $y$ was computed 
in  \cite{propagatori} (see {\bf Corollary $3.1$})
\beqa
\label{propagator}
C(x,y)&=&\int_0^\infty dt\frac{\tilde \Omega}{(2\pi {\rm sinh}\, \tilde \Omega t)^2}
e^{-\frac12
 \tilde \Omega ({\rm cotanh}\,  \tilde \Omega t )(x^2 + y^2)
-  \tilde  \Omega ({\rm cotanh}\,  \tilde \Omega t ) x\cdot y - i  \tilde \Omega x\wedge y }\nonumber\\
&=&\int_0^\infty dt\frac{\tilde \Omega}{(2\pi {\rm sinh}\, \tilde \Omega t)^2}
e^{-\frac12
 \tilde \Omega ({\rm cotanh}\,  \tilde \Omega t )(x- y)^2
- i  \tilde \Omega x\wedge y },
\eeqa
\noi
where 
 $\tom = \frac {2 \om}{\theta}$
and
\beqa
\label{def}
x\cdot y  &=& (x^1y^1+x^2y^2)+(x^3y^3+x^4y^4),\nonumber\\
x\wedge y &=& (x^1 y^2 - x^2 y^1)+(x^3 y^4 - x^4 y^3). 
\eeqa
\noi
Let us now introduce the {\it short} and {\it long variables}:
\beqa
\label{def-uv}
u=\frac{1}{\sqrt 2} (x-y),\ v=\frac{1}{\sqrt 2} (x+y).
\eeqa
\noi
Moreover let
 $\alpha_\ell= \tom t$ 
and
\beqa
\label{t}
t_\ell = {\rm tanh} \frac{\alpha_\ell}{2}.
\eeqa
The propagator \eqref{propagator} becomes
\beqa
\label{prop2}
C(x,y)=\int_0^\infty d\alpha_\ell \frac{\tilde \Omega(1-t_\ell^2)^2}{(4\pi
  t_\ell)^2}e^{-\frac 12 \tom \frac {1+ t_\ell^2}{2 t_\ell}u^2 + \tom i u\wedge v}.
\eeqa

The vertex $V$ is cyclically symmetric (note that this replaces the larger
permutational symmetry of all the fields in the vertex which holds
in ordinary commutative QFT). The vertex contribution is 
written, in position space, as (\cite{GMRV})
\beqa
\label{v1}
\delta (x_1^V - x_2^V + x_3^V - x_4^V)e^{2i\sum_{1\le
    i <j\le 4}(-1)^{i+j+1}x_i^V\Theta^{-1}x_j^V}
\eeqa
\noi
where $x_1^V,\ldots, x_4^V$ are the $4-$vectors of the positions of the $4$
fields incident to the vertex $V$.
For further use let us also define the antisymmetric matrix $\sigma$ as
\beqa
\sigma=\begin{pmatrix} \sigma_2 & 0 \\ 0 & \sigma_2 \end{pmatrix} \mbox{ with
}
\sigma_2=\begin{pmatrix} 0 & -i \\ i & 0 \end{pmatrix}.
\eeqa
\noi
The $\delta-$function appearing in the vertex contribution \eqref{v1} is
written as an integral over some new variables $p_V$, called {\it hypermomenta} \cite{GR}. 
Note that one associates such
a hypermomentum $p_V$ to any vertex $V$ {\it via} the relation
\beqa
\label{pbar1}
\delta(x_1^V -x_2^V+x_3^V-x_4^V ) &=& \int  \frac{d p'_V}{(2 \pi)^4}
e^{ip'_V(x_1^V-x_2^V+x_3^V-x_4^V)}\nonumber\\
&=&\int  \frac{d p_V}{(2 \pi)^4}
e^{p_V \sigma (x_1^V-x_2^V+x_3^V-x_4^V)}
\eeqa
\noi
where to pass from the first line to the second of the equation above one has
used the change of variable $ip'_V=p_V\sigma$, whose Jacobian is 1.

\subsection{Feynman Graphs for NCQFT}
\label{feynman}

In this subsection we give some useful conventions and  definitions. Note that this subsection is a recall of
\cite{GMRV}, \cite{V} and  \cite{propagatori}.

Let us consider a graph with $n$ vertices, $L$ internal lines and $F$
faces. 
One has
\beqa
\label{genus}
2-2g=n-L+F,
\eeqa
\noi
where $g\in\NN$ is the {\it genus} of the graph.
If $g=0$ one has a {\it planar graph}, if $g>0$ one has a {\it non-planar
  graph}. Furthermore, we call a planar graph to be a {\it planar regular
  graph} if it has no faces broken by external lines.

Such a graph has $4n$ corners, $4$ for each vertex. We denote by $N$ the
number of external positions and by $\cal I$ the set
of $4n-N$ internal corners.
The ``orientable'' form \eqref{v1} of the
vertex contribution of our model leads us to associate a sign ``+'' or ``-''
to each of the corners of each vertex. These signs alternate 
when turning around 
a vertex.
The model \eqref{actiunea} has
orientable lines in the sense of \cite{GMRV}, 
that is any internal line joins a ``-'' corner to a ``+'' corner and this is
the orientation we chose for the lines in our drawings.

Consider $ {\cal T} $ a tree of $n-1$ lines. 
The remaining $L-(n-1)$ lines
form the set $\cal L$ of loop lines.

Let us now give some ordering relations. 
If one starts from the root and turns around the tree in trigonometrical
sense, we can number each of the corners in the order they are met.

Moreover, for each vertex $V$ there is a unique tree line going towards
the root. We denote it by $\ell_V$. This correspondence works 
both ways. The {\it sign of a tree line} $\e(\ell_V)$ is
\begin{itemize}
\item -1 if the tree line is oriented towards the root and
\item 1 if not.
\end{itemize}

Vertices $V$ which are hooked to a single tree line (which is actually
$\ell_V$) are called {\it leaves}. We also define for any $\ell\in{\cal T}$ a {\it branch} $b(\ell)$ as
the subgraph containing all the vertices ``above'' $\ell$ in the tree (see
\cite{V}). 

Moreover, we define the {\it sign} $\e_k(\ell)$ {\it  of a loop line
entering/exiting a brach} associated to some vertex $k$ to be

\begin{itemize}
\item $1$ if the loop line enters the branch,
\item $-1$ if the loop line exits the branch and
\item $0$ if the loop line belongs to the branch.
\end{itemize}


Let  the lines
  $l=(i,j),\, l'=(p,q)$ and the external position $x_{k}$. We define:
\begin{itemize}
\item $  l\prec l'$ if $i<p,q$, and $j<p,q$
\item  $l\prec k$ if $i,j<k$,
\item  $l\subset l'$ if $p<i,j<q$ or $q<i,j<p$, 
\item      $k\subset l$ if $i<k<j$ or $j<k<i$,
\item      $l'\ltimes l$ if $i<p<j<q$ and $l\ltimes l'$ if $i<q<j<p$.
\end{itemize}

\begin{figure}
\centerline{\epsfig{figure=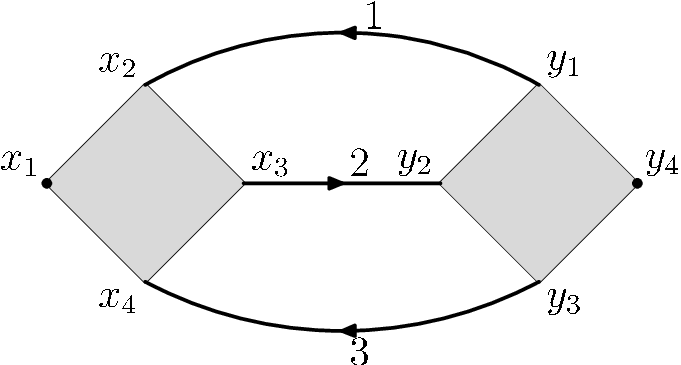,width=6cm} \hfil \epsfig{figure=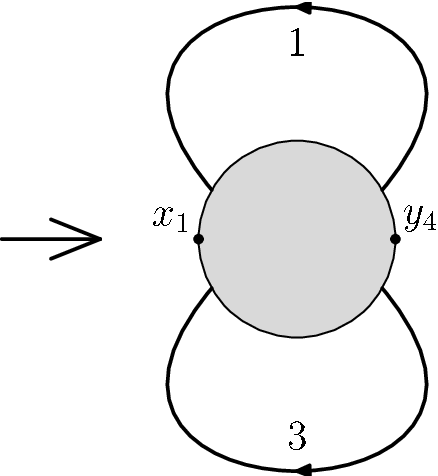,width=2cm}}
\caption{The first Filk Move: the line $2$ is reduced; the two initial
  vertices are glued up into a ``fatter'' final one}\label{firstfilk}
\end{figure}

\noindent
{\bf The first Filk move:} In \cite{filk}, T. Filk defined several contractions on a
graph or its dual which we refer to as {\it Filk moves}.
The {\it first Filk move} consists in reducing a tree line by gluing up
together two vertices into a bigger one (see Fig. \ref{firstfilk}).
Note that the number of faces or the genus of the graph do not change 
under this operation.

Repeating this operation for the $n-1$ tree lines, one obtains a single final
vertex with all the loop lines hooked to it - a {\it rosette}. If a rosette has only one
face we refer to it as to a {\it super-rosette} (see \cite{GR}).

Let us notice that the rosette can be considered as a vertex and one can write
down its vertex factor, as done for any vertex  entering some Feynman
graph. We refer to it as to the {\it rosette factor}.

Furthermore, let us remark that the order relations defined above do not change
when performing this first Filk move. Thus, as observed in section $3.1$ of
\cite{GMRV} one has

\medskip
\noindent
{{\bf Sign Alternation:}} Signs ``+'' and ``-''
alternate when turning around 
the rosette.

\medskip

We also define $ \ell <\ell' $ 
if the starting point of $\ell$ precedes the end point of $\ell'$ in the rosette.

Finally, 
once we choose a root and an orientation around the rosette,
we can define the {\it sign of a loop line} $\e(\ell_w)$ as $+1$ if
the loop line goes in the same sense as the rosette orientation and $-1$ if it
does not.

\subsection{Parametric Representation for the Noncommutative Model}

Note that, as pointed out in \cite{GR},  the first polynomial in the
noncommutative case is the determinant of the quadratic form integrated over
all internal positions of the graph {\it save one}. One has thus to chose a
particular ``root'' vertex whose position is not integrated. We denote this particular
vertex by $\bar V$.

Let us now generalize the notions \eqref{def-uv} of short and long variables
at the level of the whole Feynman graph. For this purpose we define the
$(L\times 4)$-dimensional incidence matrix $\e^V$ for each of the vertices
$V$. 
Since the graph is orientable (in the sense defined in subsection
\ref{feynman} above) we can choose
\beqa
\e_{\ell i}^V= (-1)^{i+1}, \mbox { if the line $\ell$ hooks to the vertex $V$
  at corner $i$.}
\eeqa

We also put
$$ \eta^V_{\ell,i}=\vert \e^V_{\ell,i}\vert, \mbox { } V=1,\ldots, n,\, 
\ell=1,\ldots, L \mbox{ and } i=1,\ldots, 4. $$
One now has
\beqa
v_\ell&=&\frac{1}{\sqrt{2}} \sum_V \sum_i \eta^V_{\ell i} x^V_i,\nonumber\\
u_\ell&=&\frac{1}{\sqrt{2}} \sum_V \sum_i \e^V_{\ell i} x^V_i.
\eeqa
\noi
Conversely, one has
$$ x^V_i= \frac{1}{\sqrt{2}}\left(\eta^V_{\ell i}v_\ell+\e^V_{\ell i}u_\ell \right).$$
\noi
Let us now express the amplitude $\cal A$ of such a noncommutative 
graph with the help of these long and short variables. One has to put together the expressions of all the propagators
\eqref{prop2} and vertices \eqref{v1}.  Moreover, in order to avoid the
$\sqrt{2}$ factors, we rescale the external positions $x_e$ to $\bar x_e$ and
the hypermomenta $p_V$ to $\bar p_V$. One has:
\beqa
\label{a2}
{\cal A}_G(x_e)&=& K
\int \prod_{l} {d\alpha_l}\frac{(1-t_\ell^2)^2}{t_\ell^2} \int \prod_{i \in {\cal I}}d x_i
\prod_V d p_V \  e^{-\frac{\tom}{2}\coth(\frac{\alpha_\ell}{2})u_l^2+i\tom u_\ell\wedge v_\ell}\nonumber\\
&&e^{{i}\sum_{1\le i<j\le 4}(-1)^{i+j+1}(\eta_{\ell i}^V v_\ell +
  \e^V_{\ell i} u_\ell)\Theta^{-1}(\eta_{\ell' j}^V v_{\ell'}+\e_{\ell'
  j}^Vu_{\ell' })} \  e^{\bar p_{\bar V} \sigma \sum (\eta_{\ell i}^V v_\ell+ \e_{\ell i} u_\ell)}\nonumber\\
&&e^{2i[\sum_{i \ne e}\omega(i,e)(\eta^V_{\ell i}v_\ell+\epsilon_{\ell i}^V u_\ell)
\Theta^{-1}\bar{x}_e ]+4i\sum_{e < e' } \bar{x}_e\theta^{-1}\bar{x}_{e'}+\sum_{e\in V}\bar{p}_V \sigma
(-1)^{e+1} \bar{x}_e}\ ,\nonumber \\
\eeqa
\noi
with $K$ some inessential normalization constant and $\omega(i,e)=1$ if $i<e$
and $-1$ if $i>e$.
Singling out the root vertex $\bar V$, we write
\beqa
\label{a22}
{\cal A}_G(x_e, \bar p_{\bar V})&=& K
\int \prod_{l} {d\alpha_l}\frac{(1-t_\ell^2)^2}{t_\ell^2} \int \prod_{i \in {\cal I}}d x_i
\prod_{V\ne \bar V} d p_V \  e^{-\frac{\tom}{2}\coth(\frac{\alpha_\ell}{2})u_l^2+i\tom u_\ell\wedge v_\ell}\nonumber\\
&&e^{{i}\sum_{1\le i<j\le 4}(-1)^{i+j+1}(\eta_{\ell i}^V v_\ell +
  \e^V_{\ell i} u_\ell)\Theta^{-1}(\eta_{\ell' j}^V v_{\ell'}+\e_{\ell'
  j}^Vu_{\ell' })} \  e^{\bar p_{\bar V} \sigma \sum (\eta_{\ell i}^V v_\ell+ \e_{\ell i} u_\ell)}\nonumber\\
&&e^{2i[\sum_{i \ne e}\omega(i,e)(\eta^V_{\ell i}v_\ell+\epsilon_{\ell i}^V u_\ell)
\Theta^{-1}\bar{x}_e ]+4i\sum_{e < e' } \bar{x}_e\theta^{-1}\bar{x}_{e'}+\sum_{e\in V}\bar{p}_V \sigma
(-1)^{e+1} \bar{x}_e}\ .\nonumber \\
\eeqa
\noi

From now on, in order to simplify notations, we forget the bar over
the rescaled variables $\bar x_e$ and $\bar p$, but we keep the notation
$\bar V$ for the chosen root.

One can write the amplitude \eqref{a2} in the condensed way
\beqa
\label{a-condens}
{\cal A}_G = \int \big[\frac{1-t^2}{t}\big]^{2} d\alpha \int d x d p e^{-\frac{\Omega}{2} X G X^t}
\eeqa
where 
\beqa
\label{defX}
X = \begin{pmatrix}
x_e & p_{\bar{V}} & u & v & p\\
\end{pmatrix} \ \ , \ \  G= \begin{pmatrix} M & P \\ P^{t} & Q \\
\end{pmatrix}\ .
\eeqa
Furthermore, performing the Gaussian integration one obtains:
\beqa
\label{aQ}
{\cal A}_G  = \int \big[\frac{1-t^2}{t}\big]^{2} d\alpha\frac{1}{\sqrt{{\rm det}Q}}
e^{- \frac{\tom}{2}   
\begin{pmatrix} x_e & \bar{p} \\
\end{pmatrix} [M-P Q^{-1}P^{t}]
\begin{pmatrix} x_e \\ \bar{p} \\
\end{pmatrix}  }\ .
\eeqa
\noi
This form allows to define the polynomials $HU$ and $HV$, the
noncommutative analogs 
of  $U$
and $V$ (see \eqref{s1} and resp. \eqref{s2}). 
One can write
\beqa
\label{pol}
{\cal A}_G (x_e) = K  \int_{0}^{\infty} \prod_l  [ d\alpha_l (1-t_l^2)^{2} ]
HU_G ( t )^{-2}   e^{-  \frac {HV_G ( t , x_e )}{HU_G ( t )}},
\eeqa
and resp. 
\beqa
\label{polv}
{\cal A}_{G}  (x_e,\;  p_{\bar V}) = K'  \int_{0}^{\infty} \prod_l  [ d\alpha_l (1-t_l^2)^{2} ]
HU_{G, \bar{v}} ( t )^{-2}   
e^{-  \frac {HV_{G, \bar{V}} ( t , x_e , p_{\bar V})}{HU_{G, \bar{V}} ( t )}},
\eeqa
\noi
Note that we refer to the polynomials $HU$ and $HV$ as to {\it hyperbolic
  polynomials}, since they are polynomials in the set of
  variables $t_\ell$ ($\ell =1,\ldots, L$), the hyperbolic tangent of the
  half-angle of the parameter $\alpha_\ell$ associated to each propagator line
  (see \eqref{t}). 

Using now \eqref{aQ} and \eqref{polv} the polynomial $HU_{G,
  \bar{V}}$ writes
\beqa
\label{hugvq}
HU_{G, \bar{V}}=({\rm det} \, Q)^\frac14 \prod_{\ell=1}^L t_\ell \ .
\eeqa

\section{The First Hyperbolic Polynomial}
\resetequ

We now proceed with the analysis of the polynomial $HU_{G,\bar V}$ above, the
study of the polynomial $HU_G$ being analogous. For this
purpose  we take  a closer look at  the matrix $Q$, matrix which can be read
out of the developed expression \eqref{a22}.
Note that the $(4(2L+n-1))$-dimensional  matrix $Q$ can be written as
\beqa
\label{q1}
Q=A\otimes I_{4} - B\otimes \sigma
\eeqa
\noi
where $A$ is a $(2L+n-1)$-dimensional diagonal matrix and $B$ is an
antisymmetric matrix of the same dimension.
In \cite{GR} it has been proven that for a  matrix of the form \eqref{q1}
one has
 \beqa
\label{detq}
\det Q =
[\det(A+B)]^4 .
\eeqa
\noi
which, using \eqref{hugvq} leads to
\beqa
\label{hugvq2}
HU_{G, \bar{V}}={\rm det} (A+B) \prod_{\ell=1}^L t_\ell
\eeqa
\noi
Let us now study both  the diagonal $A$ and the antisymmetric $B$ parts
of $Q$.

One has
\beqa 
\label{a}
 A=\begin{pmatrix} S & 0 & 0\\ 0  & 0 & 0 \\ 0&0&0\\
\end{pmatrix}
\eeqa 
\noi
where $S$ is a $L-$dimensional diagonal matrix with elements
$(1+t_\ell^2)\setminus(2t_\ell)$.

The antisymmetric part $B$ writes
\beqa
\label{b}
B= \begin{pmatrix}{s} E & C \\
-C^t & 0 \\
\end{pmatrix}\ 
\eeqa
\noi
with 
$$s=\frac{2}{\theta\tom}=\frac{1}{\om}$$
and 
\beqa
\label{c}
C_{\ell V}=\begin{pmatrix}
\sum_{i=1}^4(-1)^{i+1}\epsilon^V_{\ell i} \\
\sum_{i=1}^4(-1)^{i+1}\eta^V_{\ell i} \\
\end{pmatrix}\ ,
\eeqa
\noi
\beqa
\label{e}
E=\begin{pmatrix}E^{uu} & E^{uv} \\ E^{vu} & E^{vv} \\
\end{pmatrix},
\eeqa
where
\beqa
\label{ee}
E^{vv}_{\ell,\ell'}&=&\sum_V
\sum_{i,j=1}^4  (-1)^{i+j+1} \omega(i,j)\eta_{\ell i}^V\eta_{\ell' j}^V,
\nonumber\\
E^{uu}_{\ell,\ell'}&=&\sum_V
\sum_{i,j=1}^4  (-1)^{i+j+1} \omega(i,j)\epsilon_{\ell i}^V\epsilon_{\ell' j}^V,
\nonumber\\
E^{uv}_{\ell,\ell'}&=&\sum_V
\sum_{i,j=1}^4  (-1)^{i+j+1} \omega(i,j)\epsilon_{\ell i}^V\eta_{\ell'
  j}^V+ 2 \om \delta_{\ell\ell'}. 
\eeqa 
\noi
Note that $\omega$ is the antisymmetric matrix for whom $\omega(i,j)=1$ if $i<j$.

Finally in order to have the integer expression \eqref{c} of the matrix $C$ coupling
the $u$ and $v$ variables, we have rescaled by $s$ the hypermomenta $p_V$.  

We also define the integer entries matrix:
\beqa
\label{bb}
B'= \begin{pmatrix} E & C \\
-C^t & 0 \\
\end{pmatrix}\ .
\eeqa

We denote by $  |I| $
the cardinal of the set $ I$.
Let us now state the following lemma:

\begin{lemma}
\label{lemma-dev-pffaf}
With $A$ and $B$ given by (\ref{a}) and (\ref{b})
\beqa\
\label{dev-pffaf}
\det (A+B) = \sum_{\substack{I\subset \{1\dotsc L\},\\ n + |I|\; {\rm odd}}}
 (s^{-1})^{|I|+ n-1-2L}   n^2_{I}
\prod_{l\in I}\frac{1+t_\ell^2}{2t_\ell}
\eeqa
with $n_{I}=\mathrm{Pf}(B'_{\widehat{I}})$, the Pfaffian of the matrix $B'$
 above  
with {\rm{deleted}} lines and columns $I$ among the first $L$ indices 
(corresponding to short variables $u$).
\end{lemma}
{\it Proof:} The proof is straightforward, being just a particular case of
{\bf 
Lemma III.3 of \cite{GR}}. \qed

\medskip

Let us now define the integer
\beqa
\label{ki}
k_I= |I|-L-F+1.
\eeqa
\noi
Recalling that $2-2g=n-L+F$ one can use
\eqref{hugvq}, \eqref{detq} and Lemma \ref{lemma-dev-pffaf} above
to obtain  
\beqa
\label{pol-f}
HU_{G, {\bar V}} (t) &=&  \sum_{\substack{I\subset \{1\dotsc L\},\\ n + |I|\; {\rm odd}}}  s^{2g-k_{I}} \ n_{I}^2
\prod_{l \in I} \frac{1+t_\ell^2}{2t_\ell} \prod_{l' \in \{1,\ldots,L\}} t_{l'}\ .
\eeqa

\bigskip

\textbf{Leading Terms in the First Polynomial}

\bigskip

By {\it leading terms} we understand the terms of \eqref{pol-f} which have the
highest global degree in the $t_1,\ldots,t_L$ variables. It is these terms
that govern power counting. 

To obtain these leading terms one needs to express the  $I$ set in
the development \eqref{pol-f} of $HU_{G,{\bar V}}$. 
If one takes $I=\{1,\ldots , L\}$ then the corresponding Pfaffian is $0$ if
$F\ge 2$ and $2^{g}$ iff $F=1$ (by {\bf Lemma III.4 of \cite{GR}}).

Let us now consider $F\ge 2$.
We take 
\beqa
\label{i}
I=\{1,\ldots, L\}- J_0
\eeqa
\noi
where $J_0$ is an {\it admissible set} in the sense defined in \cite{GR},
{\it i.e.}
\begin{itemize}
\item it contains a tree $\tilde T$ in the dual graph and
\item its complement contains a tree $T$ in the direct graph
\end{itemize}
Then the rosette obtained by removing the lines of $J$ and contracting the lines of $T$
is a {\it super-rosette}, that is a rosette with exactly one face (see
subsection \ref{feynman}).

In \cite{GR} it was proven that
\beqa
\label{restu}
 F-1\le  |J_0| \le F-1+2g. 
\eeqa

Thus, the matrix $B'$ corresponding to \eqref{i} has an even size, which we
denote by $d$.

We take the admissible set $J_0$ such that $ |J_0|=F-1$. One then has: 
$d=(n-1)+L+(F-1)$. 

From now on, amongst the long variables $v$ we distinguish between the $n-1$ ones
corresponding to the tree lines (which we continue to call $v$) and the
$L-(n-1)$ ones corresponding to the loop lines (which we refer to as to $w$ variables).

The determinant of $B'$ writes  as a Grassmannian integral
\beqa
\label{d1}
{\rm det}B'=\int  d \bar \psi^u_1 d \psi^u_1  \ldots  d \bar
\psi^u_{F-1} d \psi^u_{F-1} d \bar \psi^w_1 d \psi^w_1\ldots  d \bar
\psi^w_{F-1} d \psi^w_{F-1}  \nonumber \\
d \bar \psi^v_1 d \psi^v_1 \ldots  d \bar \psi^v_{n-1} d \psi^v_{n-1}
d\bar \psi^p_1 d\psi^p_1\ldots d\bar \psi^p_{n-1} d\psi^p_{n-1}
 e^{-\sum_{i,j=1}^d \bar \zeta_i b'_{ij}\zeta_j}
\eeqa
\noi
where by $\zeta$ in the exponent we denote a generic Grassmannian in the set
$\{\psi^u, \psi^w, \psi^v, \psi^p\}$.
Integrating over the Grassmannian variables $\bar \psi^p_i, \psi^p_i$
($i=1,\ldots, n-1$) one gets
\beqa
\label{d2}
{\rm det}B'=\int  d \bar \psi^u_1 d \psi^u_1  \ldots  d \bar
\psi^u_{F-1} d \psi^u_{F-1} d \bar \psi^w_1 d \psi^w_1\ldots  d \bar
\psi^w_{F-1} d \psi^w_{F-1}  \nonumber \\
d \bar \psi^v_1 d \psi^v_1 \ldots  d \bar \psi^v_{n-1} d \psi^v_{n-1}
X_1^v \bar X_1^v \ldots X_{n-1}^v \bar X_{n-1}^v e^{-\sum_{i,j=1}^{(F-1)+L} \bar \psi_i b'_{ij}\psi_j}
\eeqa
\noi
with
\beqa
\label{x}
X_1^v&=& C_{1,1} \psi^u_1 + \ldots C_{L+(F-1),1}  \psi^v_{n-1} ,\nonumber\\
\bar X_1^v&=& C_{1,1} \bar \psi^u_1 + \ldots C_{L+(F-1),1} \bar \psi^v_{n-1} ,\nonumber\\
\ldots \nonumber &&\\
X_{n-1}^v&=&  C_{1,n-1} \psi^u_1 + \ldots C_{L+(F-1),n-1}  \psi^v_{n-1}
,\nonumber\\
\bar X_{n-1}^v&=&  C_{1,n-1} \bar \psi^u_1 + \ldots C_{L+(F-1),n-1} \bar \psi^v_{n-1}.
\eeqa
\noi
Note that each pair $X_i,\bar X_i$ ($i=1,\ldots, n-1$) corresponds to some
vertex $i$ (which is of course different of the root vertex).
Let us now put
\beqa
\label{c1}
\e_j\chi^v_j&=& X^v_j + \sum \e (k) X^v_k,\nonumber\\
\e_j\bar \chi^v_j&=& \bar X^v_j + \sum \e (k) \bar X^v_k
\eeqa
\noi
where the sign $\e_j$ is defined in Appendix \ref{proof} and the 
summation is performed on all the $X_k$ corresponding to the vertices
entering or exiting the branch of the vertex corresponding to $X^v_j$. 

One can now prove (see Appendix \ref{proof}) that the change of variable
\eqref{c1} is equivalent to
\beqa
\label{magic}
\psi^v_k&=&\chi^v_k - \e(k) \left[ \sum \psi^u_\ell + \sum \e_k (\ell')
  \psi^w_{\ell'} \right] ,\nonumber\\
\bar \psi^v_k&=&\bar \chi^v_k - \e_k \left[ \sum \bar \psi^u_\ell + \sum \e_k (\ell')
\bar   \psi^w_{\ell'}\right],
\eeqa
\noi
where $\e(k)$ is the sign of the tree line associated to the vertex $k$ and
$\e_k(\ell')$ is the sign of the loop line $\ell'$ which enters or exits
the branch associated to the vertex $k$ (see subsection \ref{feynman}).

Note that the form \eqref{c1} of this change of variable leads directly to
\beqa
X_1^v\bar X_1^v \ldots X^v_{n-1}\bar X^v_{n-1}=\chi_1^v\bar \chi_1^v \ldots
\chi^v_{n-1}\bar \chi^v_{n-1}. 
\eeqa
\noi
Furthermore \eqref{magic} shows that the Jacobian of this triangular change of
variables 
is $1$.

Let us remark here that the
antisymmetric character of the matrix is preserved.

The determinant \eqref{d2} rewrites as
\beqa
\label{d3}
{\rm det}\, B'
=\int \bar d \bar \psi^u_1 d \psi^u_1  \ldots \bar d \bar
\psi^u_{F-1} d \psi^u_{F-1} d \bar \psi^w_1 d \psi^w_1\ldots  d \bar
\psi^w_{F-1} d \psi^w_{F-1}  \nonumber \\
d \bar \chi^v_1 d \chi^v_1 \ldots  d \bar \chi^v_{n-1} d \chi^v_{n-1}
\chi_1 \bar \chi_1 \ldots \chi_{n-1} \bar \chi_{n-1} e^{-\sum_{i,j=1}^{(F-1)+L} \bar \zeta_i b'_{ij}\zeta_j}
\eeqa
\noi
where the general notation $\zeta$ denotes the Grassmannian variables
belonging to the set $\{\psi^u, \psi^w,\chi^v\}$ and the matrix elements
$b'_{ij}$ are obtained from the previous ones by the change of variables \eqref{magic}.

The presence of the factor $\chi_1 \bar \chi_1 \ldots \chi_{n-1} \bar
\chi_{n-1}$ in the Grassmann integral of \eqref{d3} selects only the terms with no  $\chi_1 \bar \chi_1 \ldots \chi_{n-1} \bar
\chi_{n-1}$ in  the development of the exponential. 

Thus we end up with an antisymmetric matrix of type
\beqa
\label{bprim}
B'=\begin{pmatrix}
E'^{uu}\ E'^{uw}\\
E'^{wu}\ E'^{ww}
\end{pmatrix} \ .
\eeqa

As stated above, the elements of this matrix are obtained after performing the change of
variable \eqref{magic}. This is nothing but the reduction of the graph {\it via}
the first Filk move, reduction which has as result the rosette vertex of the
graph.
One can thus just read the elements of the matrix \eqref{bprim} from the
rosette factor of the corresponding graph (see subsection \ref{feynman}) .

Let us recall that
\beqa
\label{recall}
 L= (n-1)+(F-1)+2g. 
\eeqa
\noi
This leads to a distinction between the case $g=0$ and the case $g\ge 1$ (the
non-planar case). We first treat the planar regular Feynman graphs.

\subsection{The planar regular case}
\label{planar}

Since $g=0$, the matrix $B'$ in \eqref{bprim} is
a $2(F-1)-$dimensional matrix.

Let us now give the rosette factor of a planar regular graph, a result
firstly obtained in \cite{GMRV}, presented here under
the form of 
\medskip\noindent{\bf Corollary $3.4$ of \cite{V}:}
\beqa
\label{rozeta-plan}
  \delta\big(\sum_{k=1}^{N}(-1)^{k+1}x_{k}+\sum_{l\in{\cal T}\cup{\cal L}}u_l\big)%
   \,\exp\imath\varphi\\
\eeqa
where 
\beqa
&&\varphi=\ \varphi_{E}+\varphi_{X}+\varphi_{U},\nonumber\\
    &&\varphi_{E}=\ 2\sum_{i<j,\, i,j=1}^{N}(-1)^{i+j+1}x_{i}\Theta^{-1}
    x_{j},\nonumber\\
    \nonumber\\
    &&\varphi_{X}=\ 2\sqrt{2}\sum_{k=1}^{N}
\sum_{\substack{\ell{\prec}k}}
(-1)^{k+1}x_{k}\Theta^{-1} u_{l}+ 2\sqrt{2}\sum_{\ell{\succ} k}u_{l}\Theta^{-1} (-1)^{k+1}x_{k},\nonumber\\
\eeqa
\beqa
   && \varphi_{U}=\ 2\sum\limits_{\cal T}\epsilon(l)v_{l} \Theta^{-1}u_{l}+
2\sum\limits_{\cal L}\epsilon(\ell)w_{\ell}\Theta^{-1} u_{\ell}\nonumber\\
   &&+4\sum\limits_{\ell\subset\ell', \ell'\in {\cal L}}\epsilon(\ell')w_{\ell'}\Theta^{-1}u_{l}
+4\sum\limits_{\substack{\ell\prec\ell'}}u_{l'}\Theta^{-1}
    u_{l}.\nonumber
\eeqa

Note the difference in the numerical factors with respect to 
\cite{V} or \cite{GMRV}, difference appearing from the definition \eqref{def-uv} of the
short and long variables.
 
The rosette factor \eqref{rozeta-plan} leads to the following results:

\begin{lemma}
\label{ww}
The block $E'^{ww}$ in \eqref{bprim} is identically zero.
\end{lemma}
{\it Proof:} It is straightforwrd from \eqref{rozeta-plan}. This expresses 
 the fact that in a rosette of a planar
graph the loop lines never cross each others. \qed

\begin{lemma}
\label{uw}
The lower triangular block of $E'^{uw}$ is identically zero. 
\end{lemma}
{\it Proof:} One sees from \eqref{rozeta-plan} that $ E'^{uw}_{\ell \ell'} \ne0$
if $\ell \subset \ell'$. Ordering now the loop lines one obtains the
result. \qed

\begin{lemma}
\label{uw-diag}
$$ E'^{uw}_{\ell \ell} =2 \left(\Omega - \epsilon(\ell) \right)$$
\end{lemma}
{\it Proof:} Before the reduction of the matrix {\it via} the first Filk move,
one has $E^{uw}_{\ell \ell} =2\Omega$ (coming from the propagator). The
contribution obtained from the first Filk move reduction of the tree is read
from \eqref{rozeta-plan}, thus completing the proof. \qed

\medskip

Let us also state:

\begin{lemma}
\label{pfaff}
Let $M$ be a $2d$-dimensional antisymmetric matrix such that
\begin{itemize}
\item $m_{i,j}=0,\, i=2,\ldots, d$, $d+1\le j\le d+i-1$;
\item $m_{i,j}=0,\, i=d+2,\ldots, 2d$, $d+1\le j\le i-1$.
\end{itemize}
Then
\beqa
\label{rez}
 {\rm det}M= (m_{1,d+1} m_{2, d+2},\ldots,m_{d,2d})^2.
\eeqa
\end{lemma}
{\it Proof:} 
We develop the determinant on its first line. The first
contribution to be considered is the one of $m_{1,2}$. We now develop the
remaining determinant on its first line and chose, lets say, the contribution
of $m_{2,3}$. We continue this procedure until we arrive to the contribution
of the factor $m_{d-1,d}$. From the first line of the remaining
determinant, the only non-zero element is $m_{d,2d}$. However, its
corresponding determinant is zero (the determinant of a zero-matrix).

Following the same type of arguments  one can show that the only non-zero contribution in the determinant
is the one in \eqref{rez}. \qed

\medskip

We can now put all the pieces together. By Lemmas \ref{ww}, \ref{uw} and
\ref{uw-diag} one sees that the matrix $B'$ given in \eqref{bprim} is
exactly of the form of Lemma \ref{pfaff}.
We have thus proved:

\begin{proposition}
\label{final-planar}
\beqa
{\rm Pf}\, B'= \prod\limits_{\ell\in \cal L} 2\left(\Omega - \epsilon (\ell)\right).
\eeqa
\end{proposition}

Let us now deal with the case of non-planar Feynman graphs, keeping in mind
that the interest for the planar non-regular case will be underlined in the
next section. 

\subsection{The non-planar case}

Now since $g \ge 1$ the matrix $B'$ of \eqref{bprim} is 
a $\left(2(F-1)+2g\right)$-dimensional matrix (see \eqref{recall}).
We divide the $(F-1)+2g$ loop variables $w$ in two categories:
\begin{itemize}
\item $(F-1)$ ``face'' variables, which we denote by $w^f$ and
\item $g$ pairs of ``genus'' variables, which we denote by $w^g$. 
\end{itemize}


Let us recall here the notion of ``nice-crossing'' \cite{GR} for a pair of lines of a
rosette obtained from a Feynman graph which is not regular planar. Such a pair
of lines $\ell_1$ and $\ell_2$ realize a {\it nice-crossing} if the start of
$\ell_2$ immediately precedes the end of $\ell_1$ in the rosette.

Choosing a tree in the dual graph, the respective $F-1$ lines are our
so-called face-lines (see above). The remaining rosette has only one face
({\it i.e}. it is a
super-rosette, see subsection \ref{feynman}) and hence one can state that the $g$ pairs of lines form
the $g$ nice-crossings of the rosette.

After the reduction of the graph {\it via} the first Filk move (resp. the change
of variable \eqref{magic}) the entries of the matrix $B'$ (given again by the
rosette factor, as above) are more complicated that in the planar regular
case. 

The general form of the matrix $B'$ is 
\beqa
\label{bprim-np}
B'=\begin{pmatrix}
E'^{uu}& E'^{uw^f}& E'^{uw^g}\\
E'^{w^fu}& E'^{w^fw^f}& E'^{w^fw^g}\\
E'^{w^gu}& E'^{w^gw^f}& E'^{w^gw^g}
\end{pmatrix}.
\eeqa
\noi
As before, we can read the entries of this matrix from
the rosette factor, which now writes (see {\bf Corollary $3.3$ of \cite{V}})
\beqa
\label{rozeta-np}
\delta\big(\sum_{k=1}^{N}(-1)^{j_{k}+1}s_{j_{k}}+\sum_{l\in{\cal T}\cup\cal
      L}u_l\big)%
    \,\exp\imath\varphi, 
\eeqa
\noi
with
\beqa
&&\varphi=\ \varphi_{E}+\varphi_{X}+\varphi_{U}+\varphi_{W},\nonumber\\
    &&\varphi_{E}=\ 2\sum_{k<l=1}^{N}(-1)^{j_{k}+j_{l}+1}s_{j_{k}}\Theta^{-1} s_{j_{l}},\nonumber\\
    \nonumber\\
    &&\varphi_{X}=\ 2\sqrt{2}\sum_{k=1}^{N}\sum_{\substack{\ell\prec
        j_{k}}}(-1)^{j_{k}+1}s_{j_{k}}\Theta^{-1} u_{l}
+2\sqrt{2}\sum_{\ell\succ
      j_{k}} (-1)^{j_{k}+1} u_{l} \Theta^{-1}s_{j_{k}},\nonumber\\
    \nonumber\\
    &&\varphi_{U}=\ 2\sum_{{\cal T}}\epsilon(l)v_{l}\Theta^{-1} u_{l}
    +2\sum_{{\cal L}}\epsilon(\ell)w_{\ell} \Theta^{-1}u_{\ell}\nonumber\\
    &&+2\sum_{\ell\ltimes\ell',\, \ell,\ell'\in {\cal L}}\left[\epsilon(\ell)w_{\ell}
    \Theta^{-1}u_{\ell'}+\epsilon(\ell')w_{\ell'}\Theta^{-1}
    u_{\ell}\right]
+4\sum_{\substack{\ell\subset\ell'},\, \ell'\in{\cal L}}\epsilon(\ell')w_{\ell'}
\Theta^{-1}    u_{l}\nonumber\\
    &&+4\sum_{\substack{\ell\prec\ell'}}u_{l'} \Theta^{-1}u_{l}
+2\sum_{\substack{\ell\ltimes \ell'\, \ell,\ell'\in{\cal L}}}u_{\ell'}\Theta^{-1} u_{\ell},\nonumber\\
    \nonumber\\
    &&\varphi_{W}=\ \sqrt{2}\sum_{\substack{\ell\supset
        j_{k},\, \ell\in{\cal L}}}(-1)^{j_{k}}s_{j_{k}}\Theta^{-1}\epsilon(\ell)w_{\ell}
+2\sum_{\substack{\ell\ltimes\ell',\, \ell,\ell'\in{\cal L}}}\epsilon(\ell')w_{\ell'}
\Theta^{-1}\epsilon(\ell)w_{\ell}.\nonumber
\eeqa

By a direct inspection of \eqref{rozeta-np}, one notices that 
the coupling between: 
\begin{itemize}
\item the $u_\ell$'s and the $w^f_{\ell'}$'s for whom $\ell\subset \ell'$ and
\item the $w^f$'s themselves 
\end{itemize}
is trivial. Furthermore, 
the coupling between any $u_\ell$ and its corresponding
$w^f_\ell$ is $2(\Omega-\epsilon(\ell))$. 
Moreover (as also observed in \cite{GR}), the coupling between the
$w^g$ variables has a Jordan-block form.

Nevertheless, the situation is more complicated than in the planar regular case because the new
elements $w^g$ couple in a non-trivial way with the rest of the matrix. In
order to bring these new couplings to a trivial form, we perform a sort of
``fourth Filk move'', which generalizes the third Filk move introduced in \cite{GR}.


Let us first treat the case of just one pair of
genus lines, say  $\ell^g_1\ltimes \ell^g_2$. First notice  that if some face
line crosses such a pair of genus lines, then it must cross both of the lines
of the  pair. Indeed, assuming it crosses only one of the genus lines, one can see on the rosette that
an additional face is added, which cannot be the case. 

The fourth Filk move we propose here is the following change of Grassmannian
variables: 
\beqa
\label{4filk}
&&\eta^{w^g}_1= \psi^{w^g}_1+ \sum_{\ell'<\ell_2^g;\, \ell'\cap \ell_2^g}
\psi^{w^f}_{\ell'}- \sum_{\ell''>\ell_2^g;\ell''\cap
  \ell_2^g}\psi^{w^f}_{\ell''} \nonumber\\
&& \ \ \  \ \ \  \ \ \  \ \ \  \ \ \  \ \ \  \ \ \  \ \ \  \ \ \  \ \ \ 
+ 
\sum_{\ell'<\ell_2^g;\ell'\cap \ell_2^g}
\psi^{u}_{\ell'} -
\sum_{\ell''>\ell_2^g;\ell''\cap
  \ell_2^g}\psi^{u}_{\ell''},\nonumber\\
&&\eta^{w^g}_2= \psi^{w^g}_2- \sum_{\ell'<\ell_1^g;\, \ell'\cap \ell_1^g}
\psi^{w^f}_{\ell'}+ \sum_{\ell''>\ell_1^g;\ell''\cap
  \ell_1^g}\psi^{w^f}_{\ell''} \nonumber \\
&& \ \ \  \ \ \  \ \ \  \ \ \  \ \ \  \ \ \  \ \ \  \ \ \  \ \ \  \ \ \ 
- \epsilon (\ell^g_1) \epsilon (\ell^g_2)
\left
(
- \sum_{\ell'<\ell_1^g;\ell'\cap \ell_1^g}
\psi^{u}_{\ell'} +
\sum_{\ell''>\ell_1^g;\ell''\cap
  \ell_1^g}\psi^{u}_{\ell''}
\right
),
\nonumber\\
&&\bar \eta^{w^g}_1= \bar\psi^{w^g}_1+ \sum_{\ell'<\ell_2^g;\, \ell'\cap \ell_2^g}
\bar \psi^{w^f}_{\ell'}- \sum_{\ell''>\ell_2^g;\ell''\cap
  \ell_2^g}\bar \psi^{w^f}_{\ell''} \nonumber \\
&& \ \ \  \ \ \  \ \ \  \ \ \  \ \ \  \ \ \  \ \ \  \ \ \  \ \ \  \ \ \ 
+ 
\sum_{\ell'<\ell_2^g;\ell'\cap \ell_2^g}
\bar \psi^{u}_{\ell'} -
\sum_{\ell''>\ell_2^g;\ell''\cap
  \ell_2^g}\bar \psi^{u}_{\ell''},\nonumber\\
&&\bar \eta^{w^g}_2= \bar \psi^{w^g}_2- \sum_{\ell'<\ell_1^g;\, \ell'\cap \ell_1^g}
\bar \psi^{w^f}_{\ell'}+ \sum_{\ell''>\ell_1^g;\ell''\cap
  \ell_1^g}\bar \psi^{w^f}_{\ell''}  \\
&& \ \ \  \ \ \  \ \ \  \ \ \  \ \ \  \ \ \  \ \ \  \ \ \  \ \ \  \ \ \ 
- \epsilon (\ell^g_1) \epsilon (\ell^g_2)
\left
(
- \sum_{\ell'<\ell_1^g;\ell'\cap \ell_1^g}
\bar \psi^{u}_{\ell'} +
\sum_{\ell''>\ell_1^g;\ell''\cap
  \ell_1^g}\bar \psi^{u}_{\ell''}
\right
).\nonumber
\eeqa
\noi
As proven in \cite{GR}, the terms in $\psi^w$ above will produce a
trivial coupling between all the $w$ variables (except for the coupling
between the two $w$ variables of the lines of a nice crossing coupling, which
are not affected by \eqref{4filk}; these
coupling will keep their Jordan-block form mentioned above). Furthermore,
notice that the new $\psi^u$ terms present in the fourth-Filk move
\eqref{4filk} do not affect the couplings between the $\psi^w$'s   (they just
affect the lines and columns corresponding to the $u$ variables).

Let us now investigate the change produced by \eqref{4filk} in the rest of
the couplings of the matrix. We start this investigation with the coupling
between an $u_\ell$ variable
and its corresponding $w^f_\ell$ variable. As mentioned above, before performing
\eqref{4filk} this element has the value $2(\Omega-\epsilon (\ell))$. If the
line $\ell$ does not cross the pair $\ell^g_1, \ell^g_2$ then the change of
variable \eqref{4filk} will not affect the element $E'^{uw^f}_\ell$. Suppose
now that $\ell^g_1<\ell<\ell^g_2$ (the other cases being analogous). The
effect of the $\psi^w$ terms in \eqref{4filk} on $E'^{uw^f}_{\ell, \ell}$ is the
following:
$$ E'^{uw^f}_{\ell,\ell} \to E'^{uw^f}_{\ell, \ell} + E'^{uw^g}_{u,1}+
E'^{uw^g}_{u,2}.$$
The values of $ E'^{uw^g}_{u,1}$ and 
$E'^{uw^g}_{u,2}$ can be read of \eqref{rozeta-np} to be $2\epsilon
(\ell^g_1)$ and resp. $2\epsilon
(\ell^g_2)$. One can now distinguish several cases:
\begin{itemize}
\item $\epsilon
(\ell^g_1) \epsilon
(\ell^g_2)=-1$, the contributions of the two lines cancel each other and thus
the entry $(u_\ell, w^f_\ell)$ does not change;
\item  $\epsilon
(\ell^g_1) \epsilon
(\ell^g_2)=1$, one can see from the sign alternation property (see subsection
\ref{feynman}) that if
$\epsilon(\ell^g_1)=1=\epsilon (\ell^g_2)$ then $\epsilon (\ell)=-1$ (and thus
the resp. entry changes from $2(\Omega+1)$ to $2(\Omega-1)$) or vice versa.
\end{itemize}

Now that 
this part of the change of variables \eqref{4filk} (concerning the
$\psi^{w^f}$ variables) is completed, 
the part concerning the $\psi^u$'s in \eqref{4filk} do not change
anymore the value of the $(u_\ell, w^f_\ell)$ entry. Indeed the new
contribution is now given by the entries $(w^f_\ell, w^g_1)$ and $(w^f_\ell,
w^g_2)$, which, as already stated above, are equal to $0$  after the third Filk move.

\medskip 

Using the same type of arguments one is able to prove that the coupling 
between the
$u_\ell$'s and the $w^f_{\ell'}$'s for which $\ell\subset \ell'$ remains $0$ and
furthermore that the $u$'s now couple trivially with the $w^g$'s.

\medskip

Thus, the matrix \eqref{bprim-np} has now the form 
\beqa
\begin{pmatrix}
E''^{uu} & E''^{uw^f} & 0\\
E''^{w^fu} & 0 & 0\\
0 & 0 & E''^{w^g, w^g}\\
\end{pmatrix}
\eeqa
\noi
where $E''^{w^g, w^g}$ has a Jordan-block form and $ E''^{uw^f}$ has a
lower-diagonal zero block and the elements on its  diagonal are of the form
$2(\Omega \pm 1)$. Hence the Pfaffian is $2^{g}\prod 2(\Omega \pm 1)$.

\medskip

The case of several pairs of genus-lines which present nice-crossings is
treated analogously. One performs a change of variables \eqref{4filk} for each
of the $g$ pairs of lines. The trickier cases when a face-line crosses two
pairs of genus-lines is shown by the sign alternation property (see subsection
\ref{feynman})  not to lead to a different result
than above.

\medskip

Using now \eqref{pol-f} 
we can conclude this section with:

\begin{theorem}
\label{main}
\beqa
\label{limit1}
HU_{G, {\bar V}} (t) \ge  \sum_{J_0\, {\rm admissible}}  s^{2[g+(F-1)]} \ 
\left( 2^g \prod 2 (\Omega \pm 1) \right)^2\nonumber\\
\prod_{l \in I} \frac{1+t_\ell^2}{2t_\ell} \prod_{l' \in \{1,\ldots,L\}} t_{l'}\ .
\eeqa
\end{theorem}

Let us firstly remark that the RHS above never vanishes for $\om\in[0,1[$.

As anounced in section I, this is the main result of our paper. Let us
now argue on its meaning. Indeed, using now \eqref{polv}, one has an
upper limit for the Feynman amplitude of any noncommutative graph
$G$. Moreover,  
 from 
this formula power counting follows easily, as exhibited in
\cite{GR}. Through the method proposed in this section we obtain 
an exact power
counting as function of the graph genus, hence an improvement 
 with respect to  \cite{V}. 

Furthermore we observe that the dimension $D$ of space time
 would appear simply as a parameter in this representation. 
Therefore this work
 as well as \cite{GR} is the starting point to compute dimensional
 regularization and dimensional renormalization for all classes of scalar, LSZ
 or 
 Gross-Neveu models. 

Moreover, {\it via} our approach, richer topological information is obtained,
since we deal with a positivity theorem on some ``hypertrees'' related to the
admissible sets $J_0$. The expressions deduced in this section allow
the generalization of the notion of ``democracy'' between (hyper)trees, also
present in the case of commutative QFT (see \eqref{s1}).

\medskip

Let us end this section by another remark. A weaker result than Theorem
\ref{main}  
can be obtained. 
This result does not require all the techniques used above and it just states that one has the same type of positivity 
and improved power counting
but just for a smaller class of values for
$\om\in[0,1[$, namely for the {\it transcendent} values of $\om$. 

Indeed, the Pfaffians $n_I$ corresponding to the leading terms in
\eqref{pol-f} are integer coefficient polynomials in $\om$, of degree $F-1$:
\beqa
\label{trans}
n_I=\sum_{k=0}^{F-1} a_k \om^k.
\eeqa
\noi
Developing the Pfaffian $n_I$ one can identify the
coefficient $a_{F-1}$ above as the Pfaffian $n_{I,J_0}$ of section III of
\cite{GR} (corresponding to the admissible set $J_0$). Moreover, in {\bf{Lemma
    III.5 of \cite{GR}}} it was proven that this coefficient is
nonvanishing. Therefore, the polynomial \eqref{trans} with integer
coefficients can have only a finite number of algebraic roots. These roots
could {\it a priori} vary when the graph varies, but none can be
transcendent. 

We can thus conclude that, for $\om$ a transcendent number, $n_I$ given by
\eqref{trans} never vanishes.
 Recall however that this is a weaker result than Theorem \ref{main}, which
 holds for any $\om\in[0,1[$, transcendent or not. 

\section{The Second Hyperbolic Polynomial}
\label{second}
\resetequ

We now proceed with the analysis of the second hyperbolic polynomial. As
before, we focus
on the study of $HV_{G,\bar V}$, the polynomial $HV_{G}$ being similar. This
analysis follows the same lines of  the one of section IV of \cite{GR}.

Comparing
\eqref{aQ} and \eqref{pol} one has
\beqa
\frac{HV_{G,\bar V}}{HU_{G,\bar V}}=\frac{\tom}{2}
\begin{pmatrix}x_e & p_{\bar V} \end{pmatrix}
PQ^{-1}P^t
\begin{pmatrix}x_e \\ p_{\bar V} \end{pmatrix}
\eeqa 
\noi
Note that we left aside the matrix $M$ appearing in \eqref{aQ}  since it 
 factorizes out of the integral. 

As in the previous section, the entries of the matrix $P$ are read out of
\eqref{a22}. Moreover $P$ has the same form as in \cite{GR}. 

 The expression \eqref{q1} of the matrix $Q$ implies that its inverse is:
\beqa
\label{formulinvers}
 Q^{-1}_{\tau\tau'}=\frac{(A+B)^{-1}_{\tau\tau'}+(A-B)^{-1}_{\tau\tau'}}{2}\otimes I_4+
 \frac{(A-B)^{-1}_{\tau\tau'}-(A+B)^{-1}_{\tau\tau'}}{2}\otimes \sigma.
\eeqa
\noi
One thus notices that we deal with a real part given by the $I_4$ terms and an
 imaginary part given by $\sigma$. As already mentioned in \cite{GR} we are
 interested here by the former, which leads to a hyperbolic  polynomial 
that we denote by
 $ HV^R_{G,\bar V}$.

Furthermore, let $\mathrm{Pf}(B_{\hat{K}\hat{\tau}})$ 
be  the Pfaffian of the matrix obtained from $B$ by deleting 
the lines and columns in the set ${I,\tau}$ where $\tau\notin I$. Moreover we
define $\epsilon_{I,\tau}$ to be the signature of the permutation obtained
from $(1,\ldots, d)$ by extracting the  positions belonging to $I$ and
replacing them at the end in the order
\beqa
1,\dotsc ,d\rightarrow 1,\dotsc,\hat{i_1},\dotsc ,\hat{i_p},\dotsc
,\hat{i_{\tau}},\dotsc ,d, i_{\tau},i_p\dotsc, i_1\,.
\eeqa
Note that by $d$ we mean the dimension of the matrix (here $2L+n-1$) and by
$p$ we mean the cardinal of $I$. 

One has 

\begin{lemma}
\label{lema-ultima}
\beqa
\frac{HV^R_{G,{\bar V}}}{HU_{G,{\bar V}}}(x_e, p_{\bar V})=
\frac{1}{HU_{G,{\bar V}}} \sum_I\prod_\ell t_\ell \prod_{i\in I}
\frac{1+t_i^2}{2t_i}
\Big{[}\sum_{e}x_{e}\sum_{\tau\notin K}P_{e\tau}\epsilon_{I\tau}
\mathrm{Pf}(B_{\hat{I}\hat{\tau}})\Big{]}^2.\nonumber
\eeqa 
\end{lemma}
{\it{Proof:}} The proof is straightforward, being just a particular case of {\bf
  Lemma IV.1 of \cite{GR}}. \qed

\medskip

As in the previous section, we now proceed with the analysis of the leading
terms in this sum.

\bigskip

\textbf{Leading terms}

\bigskip

As we have seen, the leading terms are given by the sets $I=\{1,\ldots,
L\}\setminus J_0$, where $J_0$ is an admissible set.

The job to be done is the investigation of the conditions under which
Pfaffians of type
\beqa
\epsilon_{I\tau}\mathrm{Pf}(B_{\hat{K}\hat{\tau}})=
\int \prod_{\alpha=1\dotsc d}d\chi_{\alpha}\prod_{i\in I}\chi_i\chi_{\tau}e^{-\frac{1}{2}\chi B\chi}\,,
\eeqa
are nonzero.


As pointed out in \cite{GR}, we first distinguish two types of graphs as
\begin{itemize}
\item the graphs which do not have any vertex with two opposite external legs and
\item the graphs which have at least a vertex with two opposite external legs.
\end{itemize}
Regarding the former case, we follow exactly the reasoning of \cite{GR}, and
for some external position $x_e$ 
we introduce a dummy Grassmann variable in
the Pfaffian integral, which once exponentiated can be interpreted as if we
add a new line between $x_e$ and the root. We denote this modified graph by
 $G'$.
One thus has a one-to-one
correspondence between the leading terms in $HV_{G,\bar V}$ and the leading
terms of $HU_{G',\bar V}$ in which the dummy line is chosen to be a tree line
(see again subsection IV.1 of \cite{GR} for further details). Note that
these leading terms were computed in section III here.

To any tree in $G'$ which contains the above dummy line there
corresponds a two-tree $T_2$ in $G$ ({\it i.e.} a tree without a line, the
dummy one). 

Let us now recall the definition of a {\it 2-admissible set} $J$ in $G$ as a set 
 which satisfies the conditions:
\begin{itemize}
\item $J$ is admissible in $G$ and
\item the dummy line is a tree of $G'$ contained in the complement of $J$.
\end{itemize}

Now, as argued in \cite{GR}, the graph obtained from $G$ by deleting the lines
$\{1,\ldots, L\}\setminus J$ and contracting the two-tree $T_2$ has two faces,
the one broken by the root and another one, which we denote by $F_J$. The bound analogous to \eqref{limit1} is
\beqa
\label{limit2}
HV^R_{G,{\bar v}}(x_e) \ge \sum_{ J \ {\rm 2-admissible \ in} \ G} 
(s^{2[g'+(F'-1)]}\left(2^{g'}\prod 2(\Omega \pm 1)\right)^2\nonumber\\
\prod_{\ell\in I}\frac{1+t_\ell^2}{2t_\ell}\prod_{\ell'}t_{\ell'}[ \sum_{e \in F_J  } (-1)^e x_e ]^2 \; .
\eeqa
\noi

The more complicated case of a graph containing at least a vertex with opposite
external legs is more tricky because, as indicated in \cite{GR} one has a sum
of two Pfaffians which could {\it a priori}  cancel each other. 
These two Pfaffians correspond to
two graphs $G_1'$ of genus $g_1$ (obtained by adding a dummy line to $G$, as
described above) and $G_2'$ (with one line erased with respect to $G_1'$) of genus
$g_2=g_1-1$. However, the second one has an additional
factor $s$ in its weight in the sum of Lemma \ref{lema-ultima} (see again
section IV of \cite{GR} for details). In our case, checking the values of the
$s$ and the $2$ factors (given by Theorem \ref{main} or \eqref{limit2}) appearing
in this sum one can easily see that no value of $\om$ can cancel it.

\section{Examples}
\label{secexamp}
\resetequ

For the  seek of completeness, we give here some examples of planar regular, non-regular and 
finally non-planar graphs.

\subsection{Planar regular graphs}

\begin{figure}[ht]
\centerline{\epsfig{figure=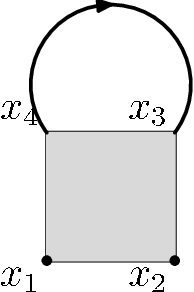,width=2cm}}
\caption{One loop, a first choice}\label{loop1}
\end{figure}

\begin{figure}[ht]
\centerline{\epsfig{figure=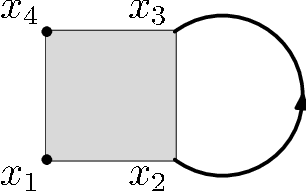,width=3cm}}
\caption{One loop, the second choice}\label{loop2}
\end{figure}

We start be the simplest example, namely the $1-$loop graph. One has here
$1$ vertex ($n=1$), one internal line ($L=1$) and two faces ($F=2$).
Applying the
methods described in this article, one finds, for the loop of Fig. \ref{loop1}
\beqa
HU_{G, \bar V}=4s^2t(\Omega +1)^2.
\eeqa
\noi
and resp.
\beqa
HU_{G, \bar V}=4s^2t(\Omega -1)^2.
\eeqa
\noi
for the loop of Fig. \ref{loop2}.

Let us now go along to a more complicated case, the bubble graph (see
Fig. \ref{figex1}) which has $n=2$, $L=2$ and $F=2$. One obtains:

\begin{figure}[ht]
\centerline{\epsfig{figure=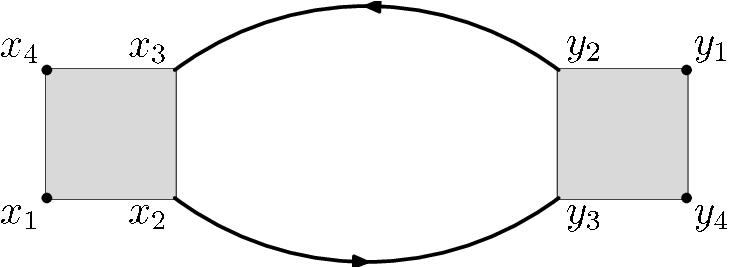,width=6cm}}
\caption{The bubble graph}\label{figex1}
\end{figure}

\beqa
HU_{G, \bar V}=2 s^2 (t_1+ t_2 + t_1^2t_2+t_1t_2^2)(\Omega -1 )^2. 
\eeqa

\begin{figure}[ht]
\centerline{\epsfig{figure=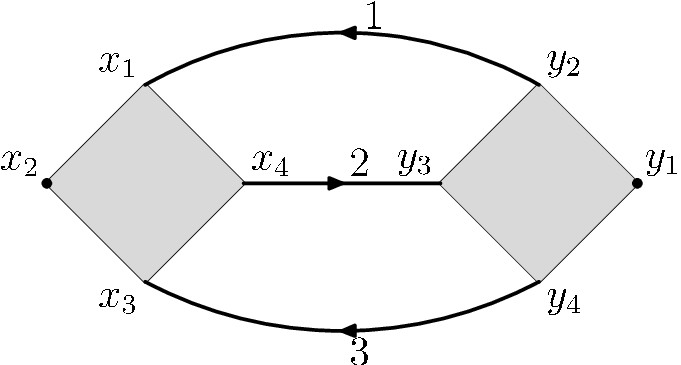,width=6cm}}
\caption{The sunshine graph}\label{figex2}
\end{figure}

For the sunshine graph (see Fig. \ref{figex2}) one has $n=2$, $L=3$, $F=3$ and

\beqa
HU_{G,\bar V}=8(1 + \Omega)^2\ s^4
\left((-1+ \Omega)^2t_2 t_3 + (-1 + \Omega)^2 t_1^2t_2 t_3 +\right.\nonumber\\ 
\left. (1 + \Omega)^2 t_1(t_2 + t_3)(1 + t_2 t_3)\right).
\eeqa

\begin{figure}[ht]
\centerline{\epsfig{figure=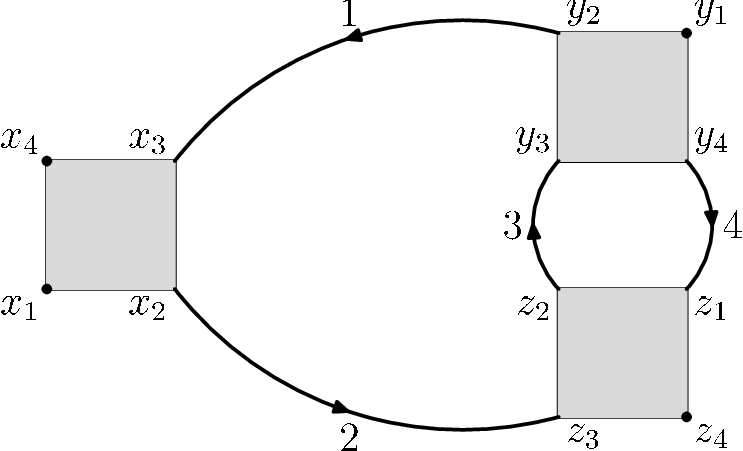,width=6cm}}
\caption{The half-eye Graph}\label{figeye}
\end{figure}

For the half-eye graph (see Fig. \ref{figeye}) one has $n=3$, $L=4$, $F=3$ and

\beqa
HU_{G, \bar V}=4 s^4 (\Omega-1)^2  (\Omega+1)^2 
[ t_3 t_4 + t_2^2 t_3 t_4 + 
        t_2 (t_3 + t_4 + t_3^2 t_4 + t_3 t_4^2) \nonumber\\
+       t_1^2\left( t_3 t_4 + t_2^2 t_3 t_4 + 
              t_2(t_3 + t_4 + t_3^2 t_4 + t_3 t_4^2)\right) \nonumber\\ 
        + t_1\left( (1 + t_2^2)(t_4 + t_3^2 t_4) + 
              t_3\left( 1 + 64 s^2 t_2 t_4 + t_4^2 + t_2^2(1 + t_4^2)\right)\right)].
\eeqa

\begin{figure}[ht]
\centerline{\epsfig{figure=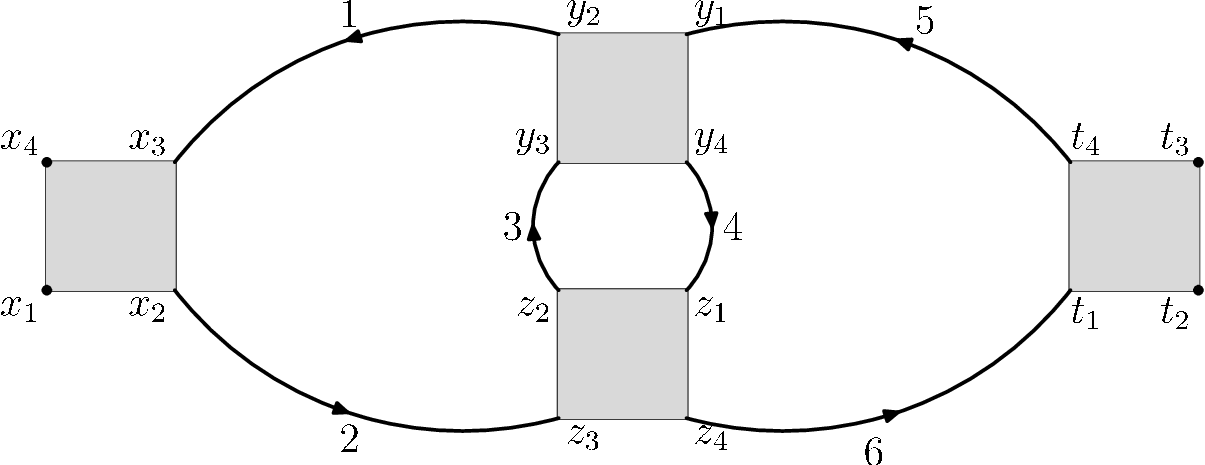,width=8cm}}
\caption{The eye graph}\label{figochi}
\end{figure}

The most complicated example of a planar regular graph we consider is the
eye-graph (see Fig. \ref{figochi}), with $n=4$, $L=6$, $F=4$. 

We obtain:
{\small{ 
\beqa
&&HU_{G, \bar V}=8\,{( -1 + \Omega ) }^4\,{( 1 + \Omega ) }^2\,s^6\,
  ( {t_3}\,{t_4}\,
     ( {t_5} + {t_6} ) \,
     ( 1 + {t_5}\,{t_6} )  \nonumber\\
&& +     {{t_2}}^2\,{t_3}\,{t_4}\,
     ( {t_5} + {t_6} ) \,
     ( 1 + {t_5}\,{t_6} )  \nonumber\\
 &&+ 
    {t_2}\,( {t_4}\,
        ( {t_5} + {t_6} ) \,
        ( 1 + {t_5}\,{t_6} )  + 
       {{t_3}}^2\,{t_4}\,
        ( {t_5} + {t_6} ) \,
        ( 1 + {t_5}\,{t_6} ) \nonumber\\
&&+       {t_3}\,( {t_5} + {t_6} + 
          {t_5}\,{t_6}\,
           ( {t_5} + {t_6} )  + 
          {{t_4}}^2\,( {t_5} + {t_6} ) \,
           ( 1 + {t_5}\,{t_6} )  \nonumber\\
&&+ 
          {t_4}\,( 1 + 64\,\Omega^2\,s^2\,{t_5}\,{t_6} + 
             {{t_6}}^2 + {{t_5}}^2\,
              ( 1 + {{t_6}}^2 )  )  )  )  \nonumber\\
&&+ 
    {t_1}\,( ( 1 + {{t_2}}^2 ) \,{t_4}\,
        ( {t_5} + {t_6} ) \,
        ( 1 + {t_5}\,{t_6} )   \nonumber\\
&&+ 
       ( 1 + {{t_2}}^2 ) \,{{t_3}}^2\,{t_4}\,
        ( {t_5} + {t_6} ) \,
        ( 1 + {t_5}\,{t_6} )  + 
       {t_3}\,( ( 1 + {{t_2}}^2 ) \,
           ( {t_5} + {t_6} ) \,
           ( 1 + {t_5}\,{t_6} )   \nonumber\\
&&+ 
          ( 1 + {{t_2}}^2 ) \,{{t_4}}^2\,
           ( {t_5} + {t_6} ) \,
           ( 1 + {t_5}\,{t_6} )  + 
          {t_4}\,( 1 + {{t_5}}^2  \nonumber\\
&&+ 
             64\,\Omega^2\,s^2\,{t_5}\,{t_6} + 
             ( 1 + {{t_5}}^2 ) \,{{t_6}}^2 + 
             64\,\Omega^2\,s^2\,{t_2}\,
              ( {t_5} + {t_6} ) \,
              ( 1 + {t_5}\,{t_6} )   \nonumber\\
&&+ 
             {{t_2}}^2\,( 1 + {{t_5}}^2 + 
                64\,\Omega^2\,s^2\,{t_5}\,{t_6} + 
                ( 1 + {{t_5}}^2 ) \,{{t_6}}^2 ) 
             )  )  )   \nonumber\\
&&+ 
    {{t_1}}^2\,( {t_3}\,{t_4}\,
        ( {t_5} + {t_6} ) \,
        ( 1 + {t_5}\,{t_6} )  + 
       {{t_2}}^2\,{t_3}\,{t_4}\,
        ( {t_5} + {t_6} ) \,
        ( 1 + {t_5}\,{t_6} )   \nonumber\\
&&+ 
       {t_2}\,( {t_4}\,
           ( {t_5} + {t_6} ) \,
           ( 1 + {t_5}\,{t_6} )  + 
          {{t_3}}^2\,{t_4}\,
           ( {t_5} + {t_6} ) \,
           ( 1 + {t_5}\,{t_6} )   \nonumber\\
&&+ 
          {t_3}\,( {t_5} + {t_6} + 
             {t_5}\,{t_6}\,
              ( {t_5} + {t_6} )  + 
             {{t_4}}^2\,( {t_5} + {t_6} ) \,
              ( 1 + {t_5}\,{t_6} )   \nonumber\\
&&+ 
             {t_4}\,( 1 + 
                64\,\Omega^2\,s^2\,{t_5}\,{t_6} + {{t_6}}^2 + 
                {{t_5}}^2\,( 1 + {{t_6}}^2 )  ) 
             )  )  )  ).
\eeqa
}}
\noi
If one chooses the admissible set to be formed of the lines $3, 4$ and $6$ the
leading term has
\beqa
n_I=8(\Omega+1)(\Omega-1)^2.
\eeqa

\subsection{An example of a planar non-regular graph}

Let us now show an example of a planar but not regular graph, the
broken bubble graph (see Fig. \ref{figex4}). This graph has $n=2$, $L=2$, $F=2$
(and thus $g=0$) but both of the faces are broken by external lines (hence
$B=2$).

One gets:

\begin{figure}[ht]
\centerline{\epsfig{figure=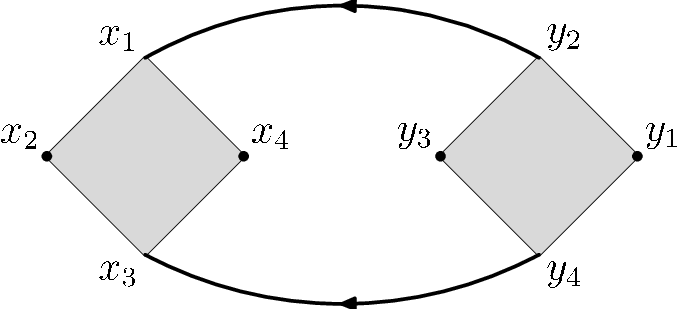,width=6cm}}
\caption{The broken bubble graph}\label{figex4}
\end{figure}

\beqa
HU_{G,\bar V}=2s^2 [ (t_2 + t_1^2 t_2) (\Omega -1 )^2 + (t_1+t_1t_2^2) (\Omega +1)^2 ].
\eeqa

\subsection{Non-planar graphs}

In order to investigate all possible cases, we conclude with some non-planar
graphs. The simplest one we consider here is the non-planar sunshine graph
(see Fig. \ref{figex3}) which has $n=2$, $L=3$ but $F=1$, and hence $g=1$. One gets:

\begin{figure}[ht]
\centerline{\epsfig{figure=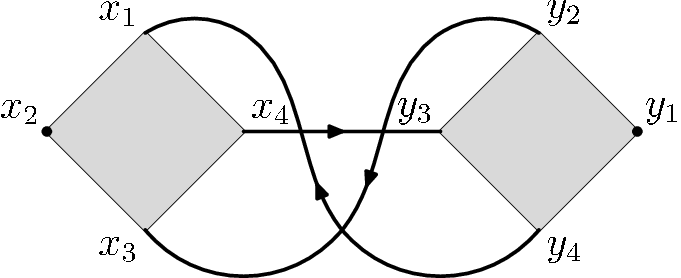,width=6cm}}
\caption{The non-planar sunshine graph}\label{figex3}
\end{figure}

\beqa
&&HU_{G, \bar V}=\frac 12 s^2(1 + 16(1 + \Omega^2)^2 s^2 t_2 t_3 + t_3^2 + 
        t_2^2(1 + t_3^2)  \nonumber\\
        &&+16(-1+2\Omega+\Omega^2)^2 s^2 t_1(t_2 + t_3 + t_2^2 t_3 + 
              t_2 t_3^2) \nonumber\\
&&+ 
        t_1^2(1 + 16(1 + \Omega^2)^2 s^2 t_2 t_3 + t_3^2 + 
              t_2^2(1 + t_3^2)).\nonumber\\
\eeqa

\begin{figure}[ht]
\centerline{\epsfig{figure=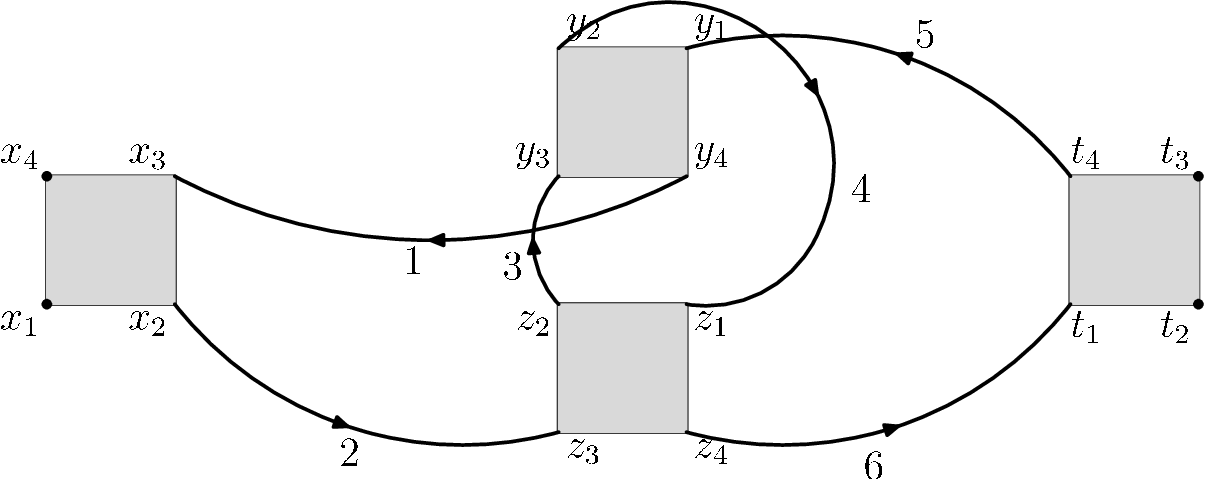,width=8.5cm}}
\caption{The twisted eye graph}\label{ochistramb}
\end{figure}

Let us consider the twisted-eye graph (see Fig. \ref{ochistramb}), which has $n=5$, $L=6$,
$F=2$ and hence $g=1$. 

If one chooses as the admissible set the set formed of the line $4$ or $6$ then one
gets
\beqa
n_I= 2^2 (\Omega -1).
\eeqa

\begin{figure}[ht]
\centerline{\epsfig{figure=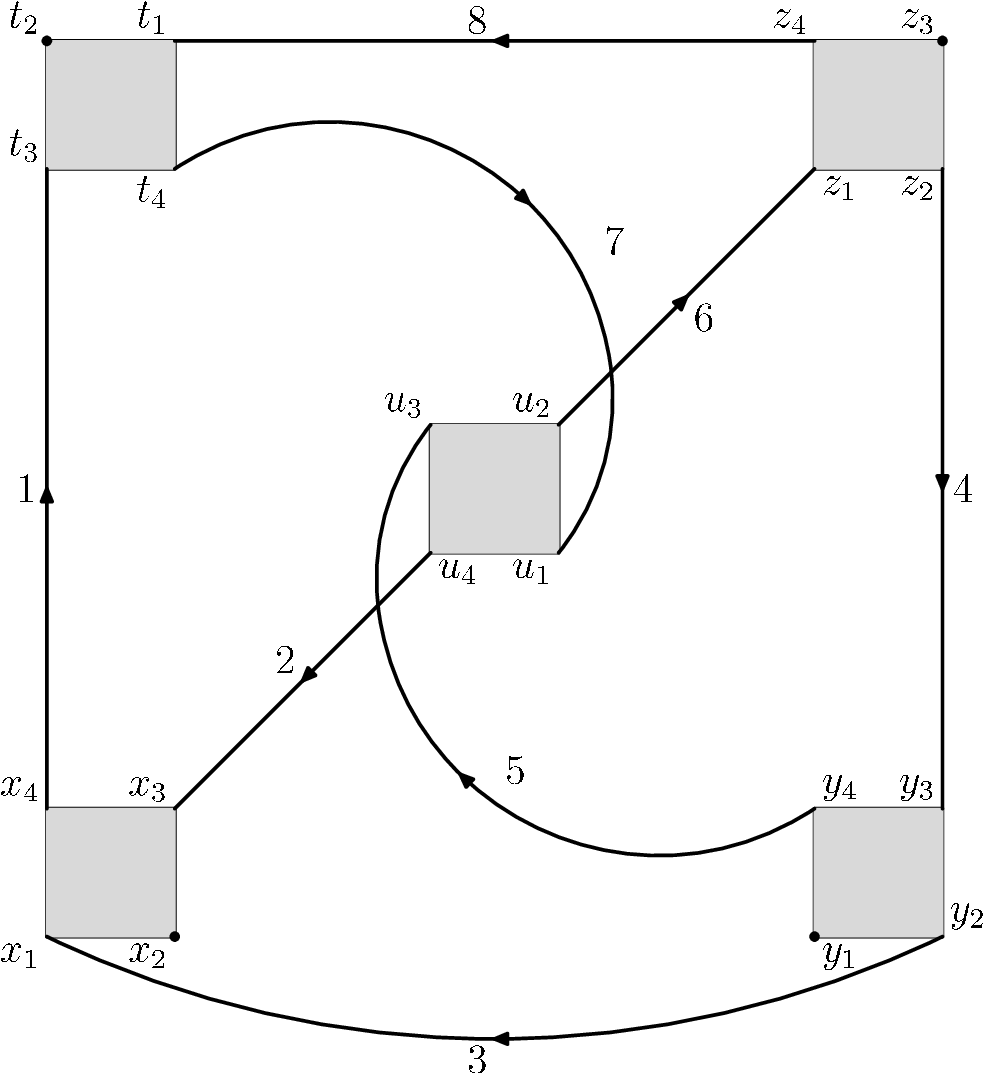,width=5.5cm}}
\caption{A more complicated non-planar graph}\label{patru}
\end{figure}

The most complicated example we have considered is the Feynman graph of
Fig. \ref{patru}. 

If one chooses the
admissible set to be the set formed of the lines $7$ and $8$, one is able to
calculate
\beqa
n_I=8(\Omega-1)(\Omega+1).
\eeqa


\appendix

\section{Proof of formula \eqref{magic}}
\label{proof}
\resetequ

We now prove that the change of variables \eqref{c1} is equivalent to the form
\eqref{magic}. We first prove that the formula is true for a
leaf vertex and then we proceed by induction to show that it is true for any
vertex in the graph.

For some vertex  $V$ one has (see \eqref{x}):
\beqa
X_V^v&=& C_{1,V} \psi^u_1 + \ldots C_{L+(F-1),V}  \psi^v_{n-1} ,\nonumber
\eeqa
\noi
When dealing  with a leaf, one has only one non-zero
entry 
$C_{i,V}$ corresponding to the set of variables
$\psi^v$. Note that this variable corresponds to the line $\ell_V$ (see
subsection \ref{feynman}). 
We denote its corresponding $\psi^v$ variable by  $\psi^v_k$ and we put, as in
\eqref{c1}
\beqa
\e_V\chi^v_V= X^v_V. \nonumber 
\eeqa
\noi
One sends the $\psi_k^v$ variable from the RHS on the LHS
and forces it to have a ``+'' sign. We now have to investigate the signs of the remaining $\psi^u$'s and
$\psi^w$'s in the RHS. Two cases are to be distinguished:
\begin{itemize}
\item the tree line $\ell_V$ exits its corresponding vertex $V$ ({\it
    i.e.} is oriented towards the root, $\epsilon (\ell_V)=-1$)
\item the tree line $\ell_V$ enters its corresponding vertex $V$ ({\it
    i.e.} is $\epsilon (\ell_V)=1$).
\end{itemize}
Consider the first of these two cases. On the RHS  one has
$C_{k,V}=-1$ and all the $\psi^u$ variables will have $+1$
coefficient. Passing $\psi^v_k$ on the LHS, one has the appropriate signs for
the $\psi^u$ of the loop lines entering the respective vertex $V$, as
indicated by formula \eqref{magic}. The case of the signs of the $\psi^w$ variables of the
second case above (when the tree line $\ell_V$ enters the vertex $V$) 
 is analogous. Finally, let us notice that the sign $\e_V$ 
is chosen such that the variable $\chi_j^v$ has a ``+'' sign on the
LHS. One can remark that this sign is nothing but the sign $\e(\ell_V)$ of the tree line $\ell_V$.
We have thus completed the proof of formula
\eqref{magic} for the case of a leaf vertex.

\medskip

We now prove by induction that our statement remains correct for any vertex in
the graph. Let us take such a general vertex $V$, with its associated tree
line going towards the root $\ell_V$. The vertex can then have a maximum of
three other tree lines connecting it to three other vertices which belong to
its branch $b(\ell_V)$. We analyze here just one of these possible other
vertices, the other ones being analyzed in exactly the same manner. Let
us call this vertex $\mu$ and thus $\ell_\mu$ is its associated line going towards the root and
joining $\mu$ and $V$. We denote its corresponding $\psi^v$ variable by  $\psi^v_j$.

As indicated in \eqref{c1}, we put
\beqa
\e_V\chi^v_V= X^v_V - \epsilon (\ell_\mu) X^v_\mu. \nonumber
\eeqa
\noi
Let us first remark that in the expression of $X^v_V$ one has $\psi^v_k$ and
$\psi^v_j$. By an analysis of the signs following the different cases of
orientations of the lines $\ell_\mu$ and $\ell_V$ one sees that the
contribution of $\psi^v_j$ cancels, thus the only $\psi^v$
variable which remains is $\psi^v_k$. Moreover, the loop present will be the
loop lines entering the two vertices, that is the loop lines entering the
branch $b(\ell_V)$. Finally, by the same type of analysis as above of the
possibilities of orientations of the lines, one concludes that the signs are the ones indicated in
\eqref{magic}.

\medskip

{\textbf{Acknowledgments}}: We thank R. Gur{\u{a}}u and F. Vignes-Tourneret for
useful discussions during the preparation of this work.

\end{document}